%% file: article_v15.tex
\documentclass[aps,pra,twocolumn,floatfix,superscriptaddress]{revtex4-1}

\usepackage{amssymb,amsmath,amstext}                %%   American Physical Society math etc extensions
\usepackage{graphicx}                                               %%   Include figure files                                            %%   Manage links
\usepackage{epstopdf}                                               %%   help with eps -> pdf 
\usepackage{color}                                                     %%   change color
\usepackage{bm}                                                        %%   bold math                                    
\usepackage{appendix}                                              %%   formating appendicies

\usepackage{bbold}
\usepackage{bbm}
\usepackage{ulem}
\usepackage{latexsym}
\usepackage[colorlinks=true,citecolor=blue,linkcolor=magenta]{hyperref}
\usepackage{cancel}
\usepackage{xcolor}
\usepackage{cleveref}
\usepackage{physics}
\usepackage{enumitem}
\usepackage{hyperref}
\usepackage{MnSymbol} %<<i,j>>
\usepackage[caption=false]{subfig}

\def\be{\begin{equation}}
\def\ee{\end{equation}}
\def\bea{\begin{eqnarray}}
\def\eea{\end{eqnarray}}

\def\bi{\begin{itemize}}
\def\ei{\end{itemize}}

\definecolor{dgreen} {RGB}{78,148,21}

\DeclareMathOperator*{\Arg}{Arg}

\DeclareMathOperator*{\Hc}{H.c.}

\begin{document} 

\title{ Determination of Chern numbers with a phase retrieval algorithm}

\author{Tomasz Szo\l{}dra} 
\affiliation{Instytut Fizyki imienia Mariana Smoluchowskiego, 
Uniwersytet Jagiello\'nski, ulica Profesora Stanis\l{}awa \L{}ojasiewicza 11, PL-30-348 Krak\'ow, Poland}
\author{Krzysztof Sacha} 
\affiliation{Instytut Fizyki imienia Mariana Smoluchowskiego, 
Uniwersytet Jagiello\'nski, ulica Profesora Stanis\l{}awa \L{}ojasiewicza 11, PL-30-348 Krak\'ow, Poland}
\affiliation{Mark Kac Complex Systems Research Center, Uniwersytet Jagiello\'nski, ulica Profesora Stanis\l{}awa \L{}ojasiewicza 11, PL-30-348 Krak\'ow, Poland
}
\author{Arkadiusz Kosior} 
\affiliation{Instytut Fizyki imienia Mariana Smoluchowskiego, 
	Uniwersytet Jagiello\'nski, ulica Profesora Stanis\l{}awa \L{}ojasiewicza 11, PL-30-348 Krak\'ow, Poland}

\date{\today}% It is always \today, today,

\begin{abstract}
Ultracold atoms in optical lattices form a clean quantum simulator platform which can be utilized to examine topological phenomena and test exotic topological materials. Here we propose an experimental scheme to measure the Chern numbers of two-dimensional multiband topological insulators with bosonic atoms. 
We show how to extract the topological invariants out of a sequence of time-of-flight images by applying a phase retrieval algorithm to matter waves. We illustrate advantages of using bosonic atoms as well as efficiency and robustness of the  method with two prominent examples: the Harper-Hofstadter model with an arbitrary commensurate magnetic flux and the Haldane model on a brick-wall lattice. 

\end{abstract}
\date{\today}

\maketitle

%%%%%%%%%%%%%%%%%%%%%%%%%%%%%%%%%%%%%%%%%%%%%%%%%%%%%%%%%%%%%%%%%%%%%%%%%%%%%%%%%%%%%%%%%%%%%%%%%%%%%

\section{Introduction}

Since Richard Feynman presented new perspectives of simulating physics  \cite{Feynman1982}, there has been an outburst of works devoted to \emph{quantum simulators} \cite{Buluta2009,Hauke2012,cirac2012goals}, which are relatively simple and controllable quantum systems that can experimentally emulate the behavior of other quantum systems or phenomena. A pronounced advantage of quantum simulators is most apparent when a targeted system is too difficult to handle for classical computers or when it is inaccessible experimentally. 

 Photonic devices \cite{aspuru2012photonic}, trapped ions  \cite{blatt2012quantum} and ultracold atoms \cite{Bloch2008,bloch2012quantum,Celi2017} are considered as the most promising quantum simulator platforms. In particular, ultracold atoms in optical lattices
constitute clean feasible systems that are free from lattice defects, phonon vibrations and electron-electron interactions. As such, these systems seem to be especially well suited to mimic miscellaneous condensed matter phenomena \cite{Jaksch2005,lewenstein:2013,Dutta2015}. By introducing fast periodic lattice modulations such as lattice shaking \cite{Eckardt2005} or laser-assisted tunneling \cite{Miyake2013} (for a review see \cite{Eckardt2017})  it is possible to study classical magnetism \cite{Struck2011,Kosior2013} and create synthetic magnetic fields for neutral atoms \cite{Jaksch2003,Kolovsky2011,Goldman2014,Goldman2014b,Celi2014} and successively design non-Abelian gauge potentials  \cite{Tagliacozzo2013nonAbelian,Kosior2014}, quantum simulators of lattice gauge theories \cite{Hauke2012b,Banerjee2012,Zohar2012,Banerjee2013,Tagliacozzo2013,Zohar2013,notarnicola2015discrete,kasper2016schwinger,Dutta2017,Zohar2017,gonzalez2017quantum} and topologically non-trivial quantum systems \cite{Miyake2013,Jotzu2014,aidelsburger:2013,kennedy2015observation,Stuhl2015,Mancini2015,goldman2016topological,kolovsky:2018}. 

The topologically protected edge conductivity in quantum Hall systems and in topological insulators is a consequence of topological properties of energy bands \cite{Halperin1982,Hasan2010,Qi2011}. As in the celebrated Harper-Hofstadter model \cite{harper:55, hofstadter:76} and the Haldane model \cite{haldane:88} (for experiments in ultracold atoms see \cite{Miyake2013,Jotzu2014,aidelsburger:2013,kennedy2015observation}), the energy bands are characterized by a non-zero value of topologically invariant Chern numbers. 
There are a few proposals how to measure the Chern numbers in a two dimensional (2D) ultracold quantum systems, including the center of mass motion \cite{Price2012,Dauphin2013,aidelsburger:2013,dauphin2017loading} and direct time-of-flight (TOF) measurements with fermionic atoms \cite{Alba2011,Hauke2014,Flaschner1091} (see also other relevant works in strip geometries \cite{Wang2013,Schweizer2016,Lu2016,mugel2017measuring} and a very recent proposal on measuring Floquet topological invariants \cite{Unal2018}).

In this paper, we propose an efficient method to determine  Chern numbers of a 2D multi-band topological insulator in a series of standard TOF measurements with a single component Bose-Einstein Condensate (BEC) prepared in an optical lattice. 
 We apply a phase retrieval algorithm \cite{Fienup:1978,Fienup1982,Fienup:86,Marchesini2007,kosior:2014} to matter waves in order to recover a small set of eigenstates that belong to the first Brillouin Zone~(BZ). We illustrate robustness of the method with two important examples:  the multiband Harper-Hofstadter model \cite{harper:55, hofstadter:76}, with an arbitrary rational flux, and the Haldane model \cite{haldane:88} on the brick-wall lattice. 

	The paper is organized as follows. In Sec.~\ref{sec:proposal} we present basic introduction to topological invariants of 2D Chern insulators and description of all elements of our method for determination of the Chern numbers. In Sec.~\ref{sec:num} we show the main results of the numerical simulations demonstrating the application of the method.  Section~\ref{sec:robustness} is devoted to an analysis of robustness of the method against experimental imperfections.
 We conclude in Sec.~\ref{sec:concl}.
	
\section{Method for determination of Chern numbers}
\label{sec:proposal}

We begin with a short introduction to Chern insulators and then we present all elements of the method for determination of Chern numbers in experiments with the help of a phase retrieval algorithm.

	\subsection{Topology of energy bands}
	\label{subsec:formalism}
	Consider a general two-dimensional tight-binding model corresponding to a square optical lattice with the lattice spacings $a_x=a_y=1$. Assume that the Hamiltonian possesses discrete translational symmetries in the configuration space: $x\rightarrow x+q$ and $y\rightarrow y+1$ where $q$ is  integer. In this case, a $q\times 1$ elementary cell has $q$ sublattice sites $\alpha = 1, \dots, q$. Due to the translation symmetry, the system shows $q$ energy bands.  An eigenstate belonging to the $n$-th band (where $n=1,\dots,q$) reads
	\begin{equation}
		\psi^{[n]}_{\bm{k}}\left(\bm{r}\right) \propto \sum_{\ell, \alpha} e^{i \bm{k} \cdot \bm{r}_{\ell \alpha}} u_{\alpha}^{[n]} \left(\bm{k}\right) w\left(\bm{r}-\bm{r}_{\ell \alpha}\right),
		\label{eq:wf}
	\end{equation}
where $w\left(\bm{r}-\bm{r}_{\ell \alpha}\right)$ is the Wannier function localized at the site $\bm{r}_{\ell \alpha}=(\alpha,\ell)$ of the optical lattice, $\bm{k}=(k_x,k_y)$  is the system quasimomentum, where	$k_x\in \left(-\pi/q, \pi/q\right]$ and $k_y \in \left(-\pi,\pi \right]$, and $u_{\alpha}^{[n]}=u_{\alpha+q}^{[n]}$  is a complex valued $q$-periodic function. Due to the translational symmetry of the system the full tight-binding Hamiltonian $\mathcal H$ can be written in a block diagonal form $\mathcal H = \bigoplus_{\bm{k}}\mathcal{H}\left(\bm{k}\right) $, where $\mathcal{H}\left(\bm{k}\right)$ are $q\times q$ blocks labeled by a quasimomentum~$\bm{k}$ \cite{bernevig:2013}. The reduced Schr\"odinger equation 
	\begin{equation}
		\mathcal{H}\left(\bm{k}\right) \bm{u}^{[n]}\left(\bm{k}\right) = E^{[n]}  \left(\bm{k}\right)\bm{u}^{[n]}  \left(\bm{k}\right),
		\label{eq:schr}
	\end{equation}
	where  $\bm{u}^{[n]} \left(\bm{k}\right) = \left[u_1^{[n]} \cdots \; u_q^{[n]}\right]^{\top}$  is the normalized eigenvector, can be solved separately for each $\bm{k}$. The eigenenergies $E^{[n]}  \left(\bm{k}\right)$ form a band.
	The geometry of energy bands can be described by
	the Berry connection $\mathcal{A}_\mu^{[n]} \left(\bm{k}\right)$ and Berry curvature $F_{xy}^{[n]}\left(\bm{k}\right)$ that read
	\begin{align}
	&\mathcal{A}^{[n]}_\mu\left(\bm{k}\right) = \bm{u}^{[n] \dagger} \left(\bm{k}\right) \partial_\mu \bm{u}^{[n]} \left(\bm{k}\right),\label{eq:berryA}\\
	&F_{xy}^{[n]}\left(\bm{k}\right) = \partial_x \mathcal{A}_y^{[n]}\left(\bm{k}\right) - \partial_y \mathcal{A}_x^{[n]}\left(\bm{k}\right),	
	\label{eq:berryF}
	\end{align}
	where $\mu = x,y$ denotes a direction in the quasimomentum space and $\partial_\mu = \partial/\partial k_\mu$ \cite{bernevig:2013}. Geometric features of energy bands can be related to topology - topological properties of the $n$-th band are characterized by the topologically invariant integer Chern number $c_n$, defined as an integral of the Berry curvature over the first BZ \cite{bernevig:2013}
	\begin{equation}
	c_n = \frac{1}{2\pi i} \int_{\text{BZ}} \mathrm{d}^2 \bm{k}~F_{xy}^{[n]} \left(\bm{k}\right), 
	\label{eq:cn}
	\end{equation} The Chern numbers determine the Hall conductance of the system if fermions are loaded to the optical lattice. The total Hall conductance is the sum of conductances of all energy bands below the Fermi level and reads 
	\be
	\sigma_{xy} = - \left(e^2 / h\right) \sum_n c_n,
	\ee
	which is the famous Thouless-Kohmoto-Nightingale-den Nijs (TKKN) formula \cite{TKKN1982}.

		In practice, it is very efficient to calculate the Chern number using the Fukui-Hatsugai-Suzuki (FHS) method \cite{FHS:2005} which, rather then a crude discretization of \eqref{eq:cn}, exploits the lattice gauge theory formalism by defining the Berry connection on the coarsely discretized BZ (see Appendix \ref{app:FHS} for details).
	
	In the following we show that applying the FHS approach and a phase retrieval algorithm \cite{Fienup:1978,Fienup1982,Fienup:86,Marchesini2007,kosior:2014} we can reconstruct Chern numbers from a series of time-of-flight experiments with a single component BEC.

\subsection{Preparation of initial eigenstates} 
	\label{subsec:prep}
	
If we knew all eigenstates of a given energy band of the Hamiltonian, then Eq.~(\ref{eq:cn}) would allow us to obtain the Chern number characterizing the band. We will show that when a BEC in the optical lattice is prepared in a certain eigenstate, measurement of the density of atoms after TOF and application of a phase retrieval algorithm allow us to reconstruct the wavefunction completely. Performing the same experiments but with a BEC in different eigenstates of the band provides sufficient information to determine the Chern number of the band. (See Sec.\ref{sec:robustness}~A for the analysis of the BZ meshing size.) In this subsection we discuss the first element of the method, i.e. the preparation of a BEC in different eigenstates of an energy band \cite{cFirst}. 

\begin{figure}
	\includegraphics[width=\columnwidth]{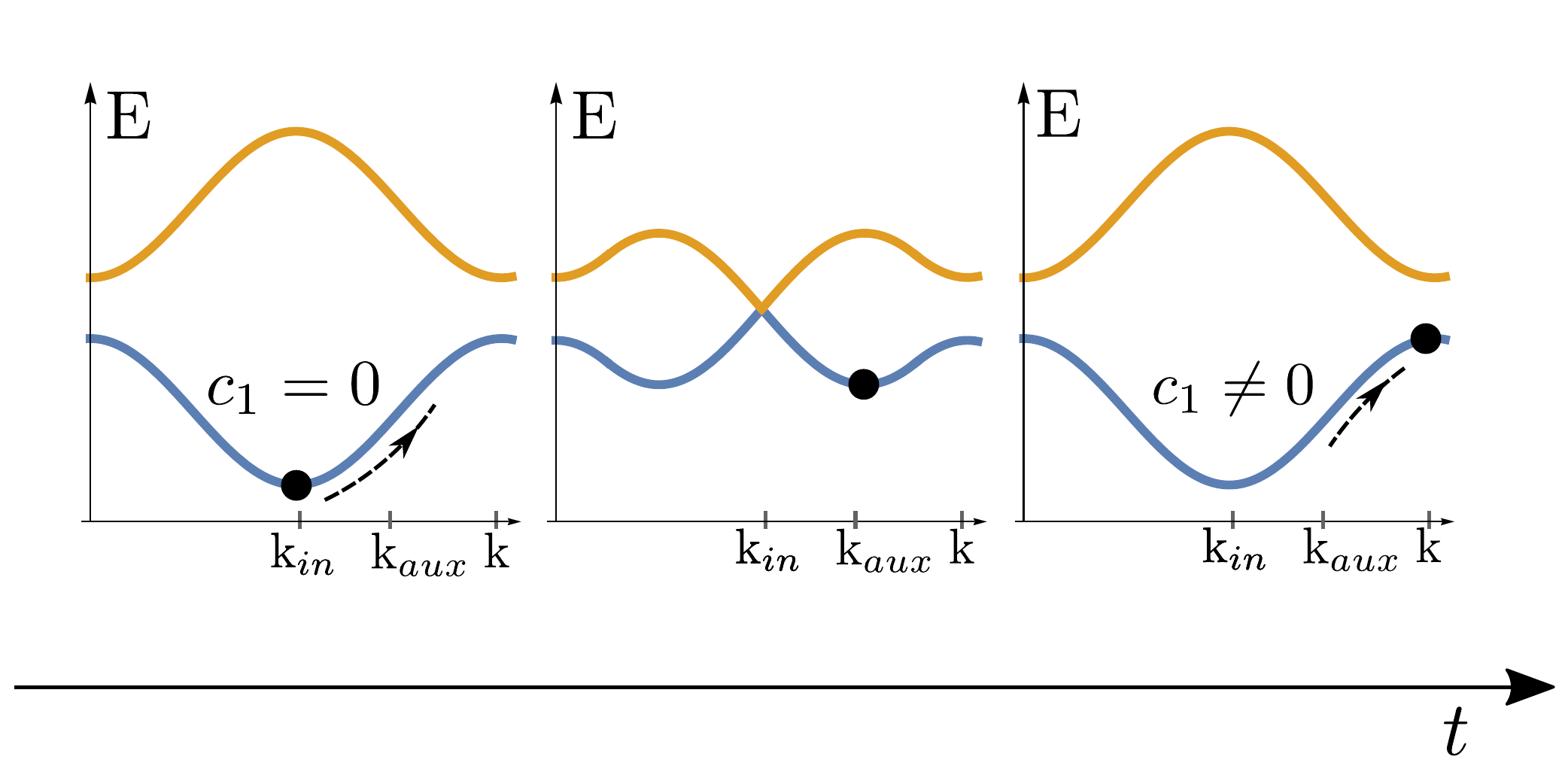}	
	\caption{Preparation of an initial eigenstate in the regime of non-trivial topology, if initially $\bm{k}_{in} \approx \bm{k}_D$. To avoid excitations to another band, before we change parameters of the system across the topological quantum phase transition, we have to apply a weak constant force $\bm{F}_1$ to shift the quasimomentum  $\bm k_{aux}=\bm k_{in} + \frac{\Delta t_1}{\hbar}\bm F_1$ far away form the Dirac point $\bm{k}_D$. At $\bm{k}_{aux}$ the upper band is not populated if the change of parameters across the topological phase transition is performed sufficiently slowly. Subsequently, i.e. after the change of system parameters to a topological regime, once again we apply a constant force to transfer a system into any final quasimomentum $\bm k= \bm k_{aux} +  \frac{\Delta t_2}{\hbar}\bm F_2$}
	\label{fig:prep}
\end{figure}

To prepare a BEC in an eigenstate corresponding to a topologically non-trivial energy band, one usually starts an experimental sequence with loading a BEC into the ground state of a 2D optical lattice with trivial topology \cite{aidelsburger:2013}. The ground state can be well-approximated by a Bloch wave~\eqref{eq:wf} with a quasimomentum $\bm k_{in}$ that minimizes the dispersion relation. By turning on  artificial gauge fields, the system is then driven into a regime of non-trivial topology of energy bands which are characterized by non-zero values of the Chern numbers~\eqref{eq:cn}. However, while switching from trivial to non-trivial topology, a quantum phase transition takes place which is accompanied by closing a gap between a neighboring band at distinct quasimomenta $\bm k_D \in \mathcal D$ (the set of Dirac points) \cite{bernevig:2013}. If $\bm k_{in}\approx \bm k_D$, in order to avoid population of another band, before we change parameters of the system across the topological quantum phase transition, we have to apply a weak constant force $\bm{F}_1$ for a suitable period of time $\Delta t_1$ so that the system is transferred to some auxiliary quasi-momentum $\bm k_{aux}=\bm k_{in} + \frac{\Delta t_1}{\hbar}\bm F_1\ne \bm k_D$ (see Fig.~\ref{fig:prep}). Then, slow change of parameters of the system across the topological phase transition does not lead to population of another band if it is done on a time scale longer than the scale given by the inverse of the energy gap corresponding to $\bm k_{aux}$. Once we are in the topological phase, we can apply another weak force $\bm{F}_2$ which allows us to transfer the system to any quasi-momentum  $\bm k= \bm k_{aux} +  \frac{\Delta t_2}{\hbar}\bm F_2$ we need. In Sec.~\ref{subsec:tof} we show how to recover full information about an eigenstate of the Bose system corresponding to a given quasi-momentum $\bm k$ in the measurement of the atomic density after TOF.
Following this experimental sequence, we can scan the whole first BZ in separate experimental realizations and obtain sufficient information about the system which allows one to determine the Chern numbers by means of the FHS approach. In the presented experimental scheme we argue that using bosonic atoms it is possible to switch to the non-trivial topology almost adiabatically by avoiding band touching points. Nevertheless, in Sec.~\ref{subsec:excitations} we present numerical studies of the influence of excitations to other bands on the determination of the Chern numbers.

	\subsection{Phase retrieval after TOF} 
	\label{subsec:tof}
	
	In this section we  review and  adapt a method \cite{kosior:2014} which allows one to reconstruct a BEC wavefunction out of a standard time-of-flight image after being processed with a phase retrieval algorithm \cite{Fienup:1978,Fienup1982,Fienup:86,Marchesini2007}. 

	A time-of-flight image shows the spatial density distribution $I\left(\bm{r}\right)$  of atoms after  a time period  $t_{TOF}$ of a free expansion  that follows a sudden  turning off an optical lattice and external trapping potentials. In the far field limit, $I\left(\bm{r}\right)$ is proportional to the initial distribution of atoms in the momentum  space if we may neglect interaction between particles during the expansion of the atomic cloud \cite{pedri2001,svistunov:2008,DEUAR2016,cSecond} 

	\begin{equation}\label{eq:intensity}
	I\left(\bm{r}\right) \propto |\tilde{\psi}_{\bm{k}} \left(\bm{q}\right)|^2 \propto \left|
	\int \mbox{d}^2\bm r e^{-i \bm q \cdot \bm r } \psi_{\bm{k}}\left(\bm{r}\right)
	\right|^2, \; \bm{q} = \frac{m\bm{r}}{\hbar t_{\text{TOF}}},
	\end{equation} 
	where $\bm{k}$ is the initial quasimomentum,  $\psi_{\bm{k}}\left(\bm{r}\right)$ and $\tilde{\psi}_{\bm{k}} \left(\bm{q}\right)$ are the representations of the initial condensate wave function in the real and reciprocal spaces, and $m$ is the atomic mass. A measurement of the atomic density reveals $|\tilde{\psi}_{\bm{k}} \left(\bm{q}\right)|^2$ at discrete points in the $\bm q$ space. If we knew not only the density but also the phase of $\tilde{\psi}_{\bm{k}} \left(\bm{q}\right)$ we would be able to obtain the wavefunction $\psi_{\bm{k}}\left(\bm{r}\right)$ by means of the inverse discrete Fourier transform. However, even without the knowledge of the phase, the task is not hopeless if we have some additional information about the system. Ultra-cold atoms are always prepared in a trap, i.e. the system always occupies finite area in the configuration space. If the  support $S$ of $\psi_{\bm{k}}\left(\bm{r}\right)$ (area where $\psi_{\bm{k}}\left(\bm{r}\right) \neq 0$) and the modulus $|\tilde{\psi}_{\bm{k}} \left(\bm{q}\right)|$ are known, one can employ an iterative phase retrieval algorithm \cite{Fienup:1978,Fienup1982,Fienup:86,Marchesini2007, kosior:2014}.  Let us stress here that the presence of an external trap is indispensable but its shape is not important as long as the trap size is significantly larger than the lattice spacing so that the quasimomentum is a good quantum number. In the present article we consider ultra-cold atoms in optical lattices and in the presence of an external hard wall potential but the phase retrieval algorithm  can be applied to other trapping potentials and lattice geometries. For example, in Ref.~\cite{kosior:2014} a 2D triangular lattice and a harmonic trapping potential are analyzed within a Thomas-Fermi approximation, where the external potential modifies the envelope of the wavefunction only.

	The phase retrieval algorithm seeks for the intersection of two sets of functions: a set of functions with a given support $S$ in the position space and a set of functions with a given modulus $|\tilde{\psi}_{\bm{k}} \left(\bm{q}\right)|$ in the reciprocal space. Let $\psi^{(i)}\left(\bm{r}\right)$ be an approximation of the desired solution at $i$-th iteration of the phase retrieval algorithm. The algorithm starts with a random, complex-valued $\psi^{(0)}\left(\bm{r}\right)$ that satisfies the support constraint $\psi^{(0)}\left(\bm{r}\right) = 0 $ for $\bm{r}~\notin~S$. In the simplest version of the algorithm \cite{Fienup:1978}, the following operations are performed at each iteration: 
	\begin{enumerate}[label=(\roman*)]
		\item The Fourier transform is performed on $\psi^{(i)}\left(\bm{r}\right)$, resulting in $|\tilde{\psi}^{(i)}\left(\bm{q}\right)| e^{i \phi^{(i)} \left(\bm{q}\right)}$.
		\label{ER:i}
		\item $|\tilde{\psi}^{(i)}\left(\bm{q}\right)|$ is substituted with the true $|\tilde{\psi}_{\bm{k}} \left(\bm{q}\right)|$ which is obtained in an experiment after TOF.
		\label{ER:ii}
		\item Inverse Fourier transform is applied which gives $\psi^{(i+1)}\left(\bm{r}\right)$, not necessarily satisfying the support constraint.
		\label{ER:iii}
		\item The support constraint is imposed on $\psi^{(i+1)}\left(\bm{r}\right)$ by setting $\psi^{(i+1)}\left(\bm{r}\right)=0$ for every $\bm{r} \notin S$.
		\label{ER:iv}
	\end{enumerate}
Convergence of the algorithm is tracked by the error measure defined as
\be
\varepsilon = \int \mathrm{d}^2 \bm{q}~\left( \left|\tilde{\psi}^{\text{ ret}}\left(\bm{q}\right)\right| - \left|\tilde{\psi}_{\bm{k}}\left(\bm{q}\right)\right| \right)^2, 
\label{epsilon_error}
\ee
where $\tilde{\psi}^{\text{ret}}_{\bm{k}}\left(\bm{q}\right)$ is a retrieved function and  $\left|\tilde{\psi}_{\bm{k}}\left(\bm{q}\right)\right|^2 $ is the measured  probability distribution.
	The presented simplest version of the algorithm guarantees a decrease of $\varepsilon$ in each iteration. Unfortunately, once it reaches a local minimum of $\varepsilon$, it cannot proceed further. There are modifications of the phase retrieval methods which allow for the much faster convergence to a desired solution \cite{Fienup1982,Fienup:86,Marchesini2007}.  Moreover, to increase the rate of the convergence one can use any extra information about $\psi_{\bm{k}}\left(\bm{r}\right)$, e.g., a preliminary \emph{in-situ} measurement of $\left|\psi_{\bm{k}}\left(\bm{r}\right)\right|$ or its theoretical estimation \cite{kosior:2014}. In our case, we speed up the convergence by exploiting information about geometry of an optical lattice only, i.e. we do not assume anything about the parameters of the Hamiltonian, see Appendix \ref{app:PR} for all details. 
	
	Once $\psi_{\bm{k}}\left(\bm{r}\right)$, Eq.~\eqref{eq:wf}, is successfully recovered, in order to extract the coefficient vector $\bm{u} \left(\bm{k}\right)$ one has to project $\psi_{\bm{k}}\left(\bm{r}\right)$ on the orthonormal basis of the Wannier functions. To minimize the numerical error one might additionally average each $u_\alpha$ component over lattice sites $\ell = 1, \ldots, n_{\text{cells}}$:
		\begin{equation}
	u_\alpha \left(\bm{k}\right) = \frac{1}{n_{\text{cells}}} \sum_{\ell} e^{-i \bm{k}\cdot \bm{r}_{\ell \alpha}} \int \mathrm{d}^2\bm{r}~w^*\left(\bm{r} - \bm{r}_{\ell \alpha}\right)\psi_{\bm{k}}\left(\bm{r}\right).
	\label{eq:usRetr}
	\end{equation}
The Wannier functions $w(\bm{r})$ can be well approximated by Gaussian functions with the width $\sigma_{w}= \hbar t_{\text{TOF}}/\left(m \sigma_{\tilde w}\right)$  which can be obtained from  the wide envelope of the measured density profile 
\be
 |\tilde{\psi}_{\bm{k}}(\bm q)|^2 \propto |\tilde{w}\left(\bm{q}\right)|^2\left|
 \sum_{\ell, \alpha} e^{i (\bm{k} -\bm q) \cdot \bm{r}_{\ell \alpha}} u_{\alpha} \left(\bm{k}\right)
 \right|^2,
\ee
where  $\tilde{w}\left(\bm{q}\right)$ is the Fourier transform of $w(\bm r)$ and $\sigma_{\tilde{w}}$ is the width of $\tilde{w}\left(\bm{q}\right)$.
	
	\subsection{Calculation of the Chern number}
	\label{subsec:chern}
      In order to determine the Chern number we propose a series of experiments with a BEC in an optical lattice. In each experiment, one prepares a BEC in an eigenstate with a different quasimomentum $\bm{k}$ from the first BZ and retrieves a column complex-valued vector $\bm{u}\left(\bm{k}\right)$, Eq.~\eqref{eq:schr},  using the phase retrieval algorithms (see Sec. \ref{subsec:tof}). 
      To obtain the Chern number we apply a highly effective FHS method \cite{FHS:2005} which allows us to calculate the Chern number with the help of a few eigenvectors $\psi_{\bm k}(\bm r)$ only, i.e. the coarsely discretized BZ. It is possible due to the fact that the FHS algorithm is based on a gauge-invariant lattice gauge theory formulation.  (See Appendix \ref{app:FHS} for a quick
	revision of the FHS algorithm.) 
	In Sec. \ref{sec:num} we demonstrate the method of the determination of the Chern numbers simulating experimental data for two examples: Harper-Hofstadter and Haldane models.

\section{Numerical simulations}
\label{sec:num}
The proposed experimental scheme of detecting Chern numbers applies to a general tight-binding  Hamiltonian in a two-dimensional space. In this section we illustrate application of the scheme with two examples: the Harper-Hofstadter model with an arbitrary rational flux \cite{harper:55,hofstadter:76} and the Haldane model \cite{haldane:88} on a brick-wall lattice (for experiments in ultracold atoms see \cite{Miyake2013,Jotzu2014,aidelsburger:2013,kennedy2015observation}). In the case of the Harper-Hofstadter model we show that a large number of bands is not the limitation of our method. With the help of the Haldane model we demonstrate that our scheme allows one to reconstruct the phase diagram of the system.

	\subsection{The Harper-Hofstadter model}
	\label{sec:hh}
	\begin{figure}		 		
		\includegraphics[width=\columnwidth]{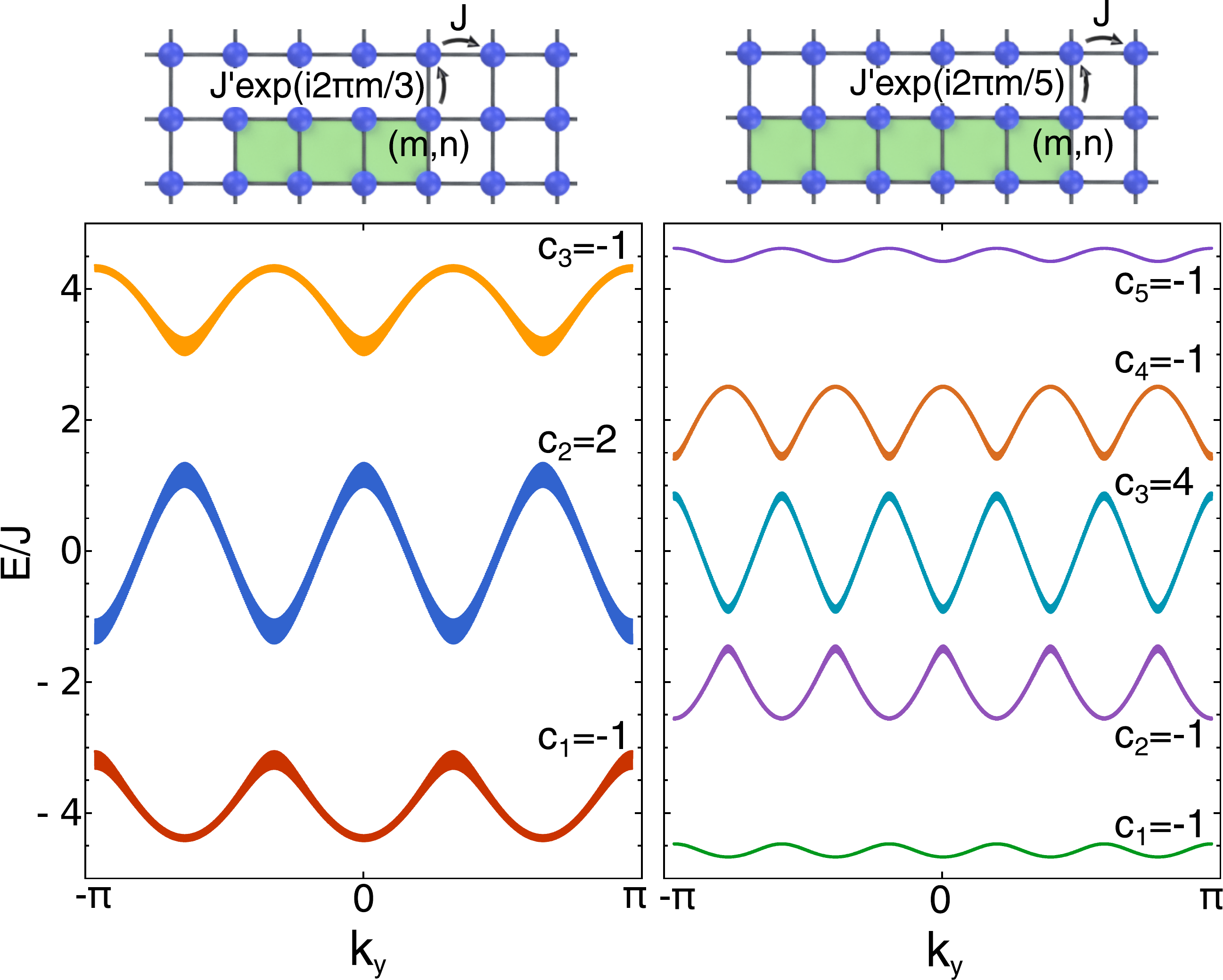}       
		 		\caption{Harper-Hofstadter model is a 2D square lattice with tunneling amplitudes $J$, $J'$, in $x$, $y$, pierced by uniform artificial magnetic field. A particle traveling along $y$ acquires the Peierls phase. We denote (magnetic) elementary cells by green rectangles for two magnetic fluxes through plaquette, $\phi=1/3$ (left panel) and $\phi = 1/5$ (right panel). The corresponding energy spectra are calculated for $J/J' = 1/2$. Chern numbers associated to energy bands are indicated.  }
		 		\label{fig:hofstadter} 
 	\end{figure} 

 Consider bosonic atoms in a square two-dimensional lattice, in $XY$-plane, with a unit lattice spacing   subjected to uniform artificial magnetic field $\bm{B} = \left(0,0,B\right)$. The nearest-neighbor-hopping Hamiltonian of an atom in the Landau gauge with the vector potential $\bm{A} = (0, Bx, 0)$ takes the following form
	\begin{equation}
	\hat{\mathcal{H}} = -\sum_{m,n}\left( J \hat{c}^\dagger_{m+1, n} \hat{c}_{m,n} + J' e^{i 2\pi \phi m} \hat{c}^\dagger_{m, n+1} \hat{c}_{m,n} + \Hc\right),
	\label{eq:hhHamiltonian}
	\end{equation}
	where $\hat{c}^\dagger_{m,n}, \hat{c}_{m,n}$ are the bosonic particle creation and annihilation operators corresponding to a lattice site $(m, n)$. $J,\;J'$ are tunneling amplitudes and $\phi = B/h$ is a dimensionless flux. Due to the presence of the magnetic field, particles tunneling along $y$ acquire the Peierls phase factor $e^{i 2\pi \phi m}$ \cite{peierls:33}. The presence of the magnetic field, in principle, breaks discrete space-translation symmetry of the lattice.  However, if the flux is a rational number, $\phi = p/q$ where $p$ and $q$ are coprime integers, the translational symmetry is restored but with the spatial period $q$ times longer than the lattice constant.  Therefore, an effective magnetic $q\times 1$ elementary cell consists of $q$ lattice sites, and the first BZ is the rectangle $2\pi/q \times 2\pi$ in the quasi-momentum space.  After rewriting the Hamiltonian \eqref{eq:hhHamiltonian} in the Fourier space, the reduced Schr\"odinger equation \eqref{eq:schr} takes the following form:
	\begin{align}\label{eq:harper}
	-J e^{i k_x } u_{\alpha+1}\left(\bm{k}\right)- 2 J' &\cos\left(k_y  +2\pi\frac{p}{q}\alpha\right) u_\alpha \left(\bm{k}\right)  \\\nonumber
	&- J e^{-i k_x  } u_{\alpha-1} \left(\bm{k}\right) = E\left(\bm{k}\right) u_\alpha \left(\bm{k}\right),
	\end{align}
 where $\alpha=1,\ldots,q$.

 \begin{figure}[bt]
 \includegraphics[width=.9\columnwidth]{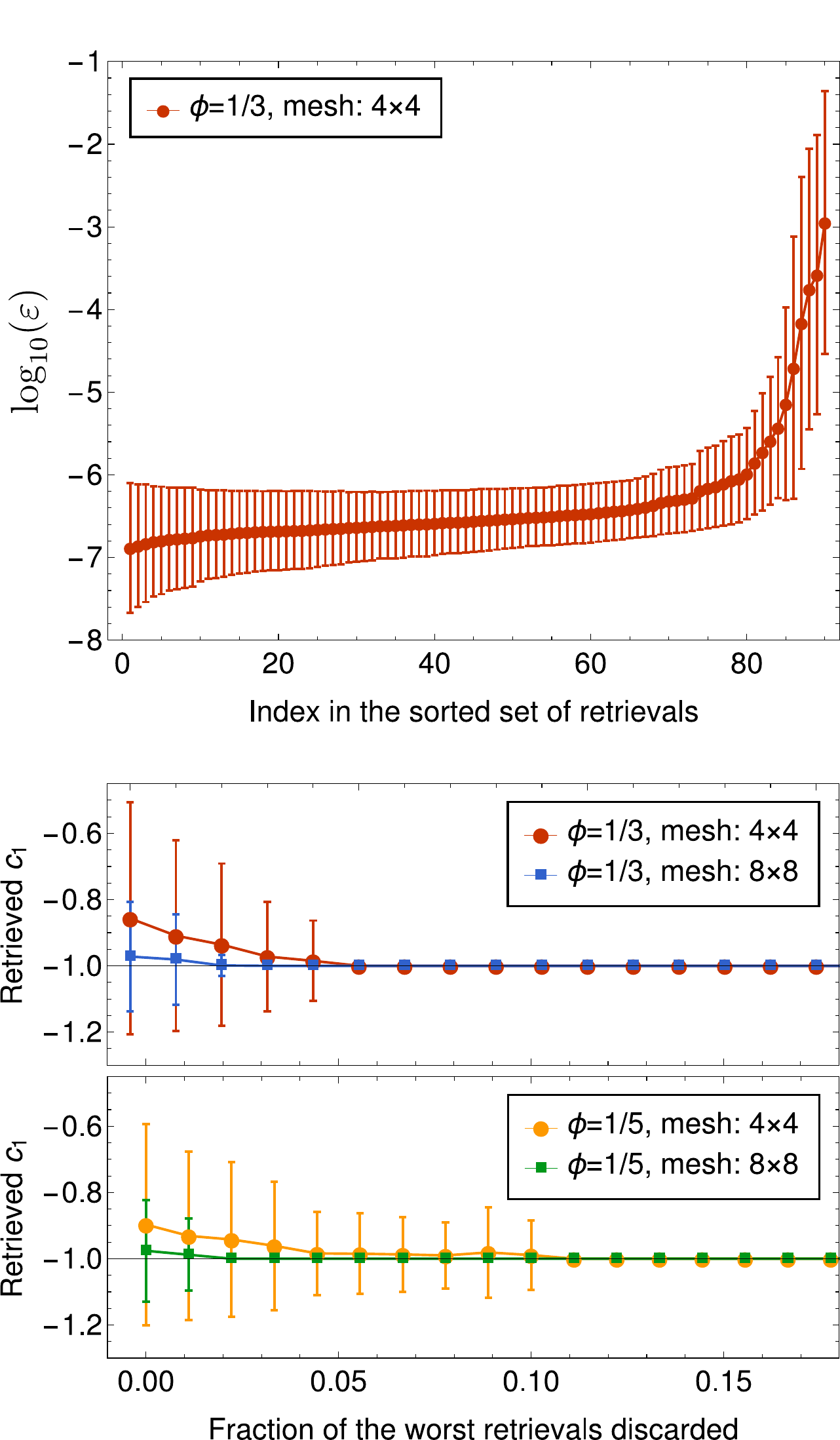}
	\caption{Reconstruction of the Chern number for the lowest band of the Harper-Hofstadter model with magnetic flux $\phi = 1/3$ and $\phi=1/5$. Upper panel: Mean (over distinct quasimomenta) of logarithm of sorted retrieval errors $\varepsilon$, see Eq. \eqref{epsilon_error}. On average, about 90 \% of independent phase retrieval runs converge successfully $\left(\varepsilon \approx 10^{-6}-10^{-7}\right)$. Lower panel: The reconstructed Chern numbers as a function of a percentage of rejected retrievals. The calculations give the proper value $c_1=-1$ within error bars even if all unsuccessful retrievals are selected. Although the better results are obtained for a higher mesh size ($8 \times 8$), after rejecting about 10\% of the worst retrievals, a very coarse mesh ($4\times 4$) already gives a perfect agreement with the model.}
	\label{fig:pr}
\end{figure}

Let us focus on the reconstruction of the lowest band Chern number for $q = 3$ and $q=5$ band models, as depicted   in Fig.~\ref{fig:hofstadter}. We choose a finite optical lattice consisting of $7\times 7$ effective magnetic elementary cells (which corresponds to $7\times 21$ or $7\times 35$ lattice sites for $q = 3$ and $q=5$ bands respectively). In principle, the measurement of $|\tilde\psi_{\bm k}(\bm q)|$ and performing the phase retrieval algorithm allows us to obtain the full information about the eigenstate $\psi_{\bm k}(\bm r)$ and successively recover the Chern number (see Sec.~\ref{subsec:tof}~-~\ref{subsec:chern}). However, the phase retrieval algorithm is known to occasionally get stuck at local minima. Therefore, for every $\bm{k}$ we repeat the algorithm, each time starting from different randomly generated initial state. 

 All retrieved eigenstates can be sorted by their error $\varepsilon$, Eq. \eqref{epsilon_error}, as shown in Fig.~\ref{fig:pr}(upper panel). 
 It is evident that about the 90\% of the best phase retrieval runs converge to functions with approximately the same error $\varepsilon \approx 10^{-6}-10^{-7}$, while the errors of the last 5-10\% trails are larger by a few orders of magnitude. 
 
 For each quasimomentum $\bm k$ we select a random representative out of 90 phase retrieval algorithm runs and calculate the Chern number with the FHS method. We repeat the process $10^3$ times and successively average the data. (Let us stress that this repetition is a data processing post measurement only.) As we illustrate in~Fig.~\ref{fig:pr}(lower panel), after rejecting the worst phase retrieval trails we are always able to recover the Chern numbers $c_1=-1$ with a perfect accuracy. Note that without any rejections, for a 8$\times$8 BZ mesh we  obtain $c_1=-0.97(17)$ for $\phi=1/5$ and $c_1=-0.98(15)$ for $\phi=1/3$. Moreover, in  Fig.~\ref{fig:pr} we show that a much harsher discretization of the first BZ (4$\times$4 mesh is already sufficient to correctly recover the Chern number.

	\subsection{Haldane model on a brick-wall lattice}
	\label{sec:haldane}
	 	\begin{figure}[bt] 	 
		
		\includegraphics[width=\columnwidth]{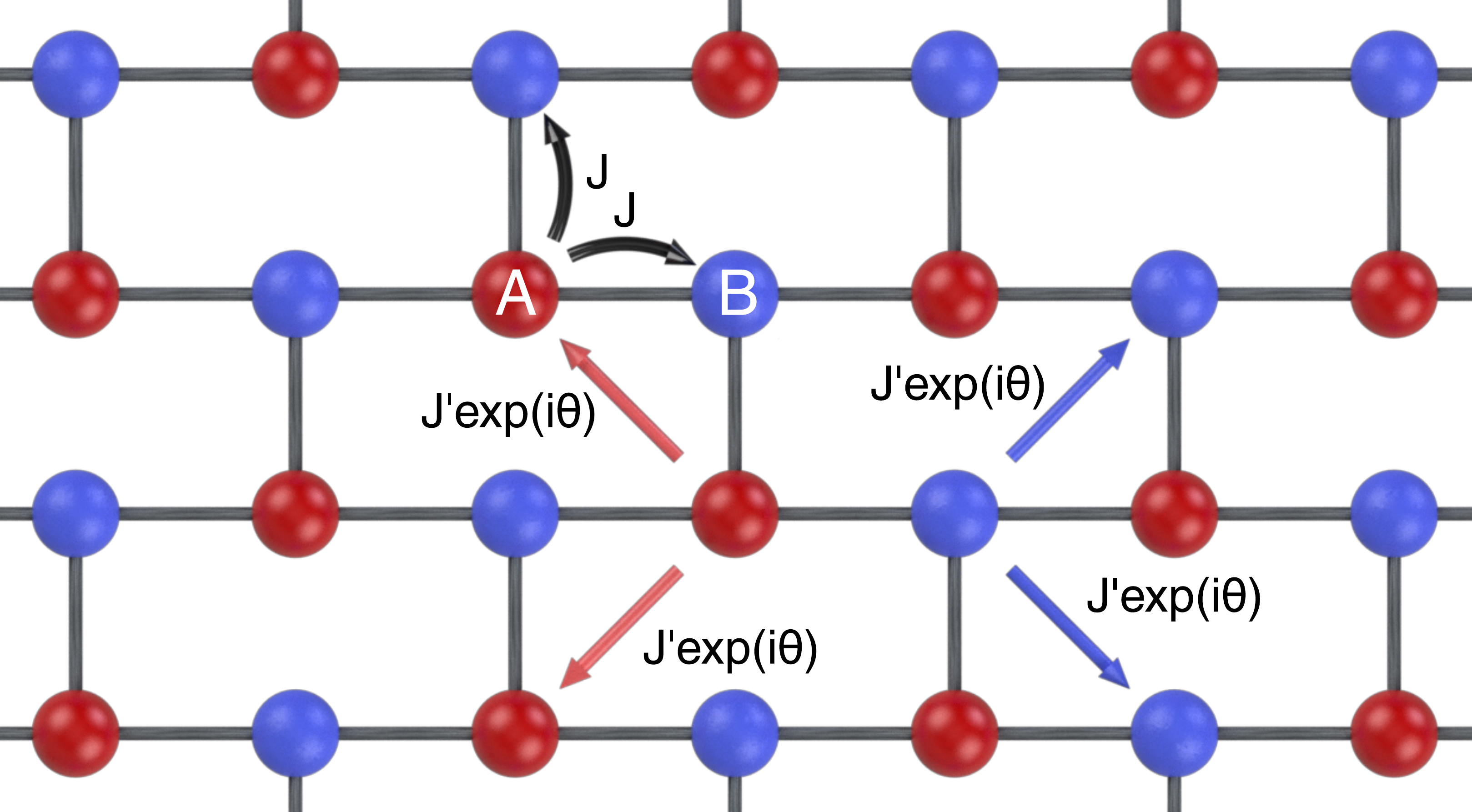} 
		\caption{Haldane model on a brick-wall lattice -- two interpenetrating square lattices, with real and complex tunnelings to the nearest and next-nearest neighboring sites, respectively. Arrows denote directions of the tunnelings.}
		\label{fig:haldane} 
	\end{figure} 

	The brick-wall structure consists of two interpenetrating square lattices $A$ and $B$, see Fig.~\ref{fig:haldane}. We assume real tunneling amplitudes $J$ between nearest neighboring lattice sites and complex tunneling amplitudes $J' e^{\pm i\theta}$ between next-nearest neighboring sites.  The model is topologically equivalent to the Haldane model on a honeycomb lattice \cite{haldane:88}. The Hamiltonian of the system reads
	\begin{equation}
		\hat{\mathcal{H}} = -J \sum_{\left\langle i,j \right\rangle} \hat{c}_i^\dagger \hat{c}_j - J' \sum_{\llangle i,j \rrangle} e^{i \theta_{ij}} \hat{c}_i^\dagger \hat{c}_j + \Delta \sum_i \epsilon_i \hat{c}_i^\dagger \hat{c}_i,
	\end{equation}
	where $i,j$ are indices of the lattice sites, $\left\langle i, j \right\rangle$ denotes pairs of nearest neighbors, $\left\llangle i, j\right\rrangle$ pairs of next-nearest neighbors, $\theta_{ij} = \pm \theta$ where the sign depends  on the direction of the tunneling, $\Delta$ introduces the energy offset between the $A$ and $B$ sublattices because $\epsilon_i = 1$ for $i\in A$,  $\epsilon_i = - 1$ for $i \in B$ (see Fig.~\ref{fig:haldane}). Complex values of the tunneling amplitudes break the time-reversal symmetry while the energy offset breaks the parity symmetry. Switching to the reciprocal space we can write the Hamiltonian in a block diagonal form. Each block is a $2\times 2$ matrix $\mathcal{H}(\bm k)$ whose elements  take the form

	\begin{align}
				&\mathcal{H}_{11} = \Delta- 2 J' (\cos\left(\theta + 2 k_x \right) + \cos \left(\theta - k_x  - k_y \right) \nonumber \\ \nonumber
				&\qquad\quad+ \cos \left(\theta - k_x  + k_y \right)),\\\nonumber
				&\mathcal{H}_{12} = \mathcal{H}_{21}^* = - J (2 \cos k_x  + e^{-i k_y }),\\\nonumber
				&\mathcal{H}_{22} = -\Delta - 2 J' (\cos\left(\theta - 2 k_x \right) + \cos \left(\theta + k_x  + k_y \right) \\\nonumber
				&\qquad\quad+ \cos \left(\theta + k_x - k_y \right)).
	\end{align}

\begin{figure}[bt] 	 
		
		\includegraphics[width=\columnwidth]{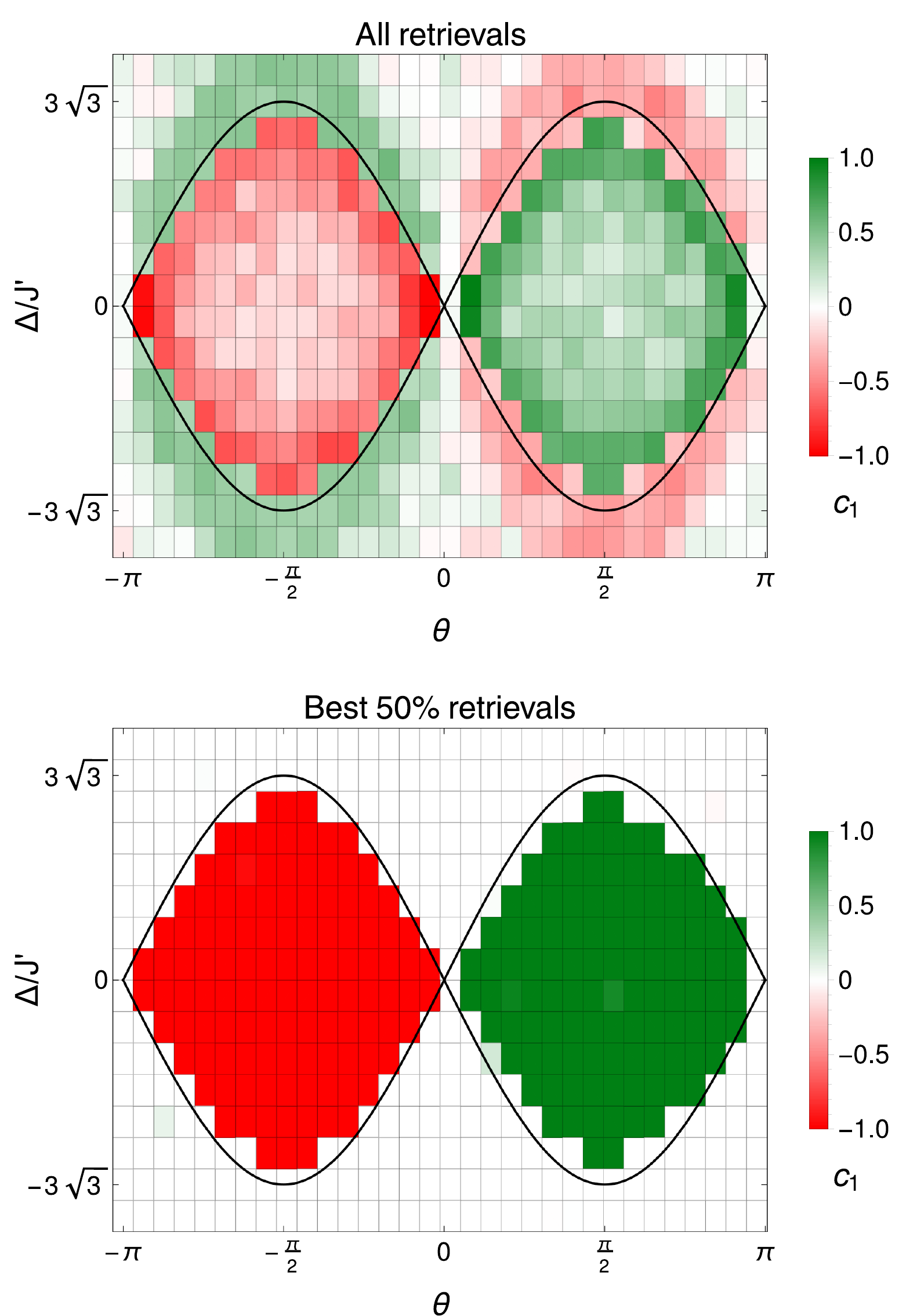}
		\caption{Topological phase diagram of the lowest band in the Haldane model, obtained from simulated TOF images using phase retrieval algorithm. Black lines indicate phase transitions at $\pm 3\sqrt{3} \sin \theta$, predicted by Haldane \cite{haldane:88}. The quality of the phase diagram depends on the percentage of rejected phase retrieval outputs. Upper panel: When all the phase retrieval runs are taken into account, only general features of the phase diagram are reproduced. Lower panel: Rejection of 50\% worst results (according to the retrieval error $\varepsilon$), already leads to an exact reconstruction of the phase diagram and more rejections do not change the picture. }
		\label{fig:phasediagram} 
	\end{figure}

An identical procedure as in the case of the Harper-Hofstadter model leads to a successful retrieval of the Chern number of the lowest band. This allows us to obtain the topological phase diagram of the Haldane model, see Fig.~\ref{fig:phasediagram}. The discretization of the first BZ corresponds to the $6\times6$ mesh. For each of the eigenstates we assume that we know the support of $\psi_{\bm k}(\bm r)$ and the modulus $|\tilde\psi_{\bm k}(\bm q)|$ and perform the phase retrieval procedure 90 times. Each application of the algorithm starts with randomly chosen phases of an eigenstate and consists of 350 iterations. We may now select a number of the best results, based on their error $\varepsilon$, Eq. \eqref{epsilon_error}, and make statistics on the retrieved Chern numbers, as in Sec. \ref{sec:hh}. Taking all results, including those that did not converge to a global solution, we obtain a topological phase diagram in Fig. \ref{fig:phasediagram}~(upper panel) which only qualitatively represents a structure predicted by Haldane \cite{haldane:88}. However, selecting 50\% of the best results yields a perfect recovery of the Haldane model phase diagram, shown in Fig.\ref{fig:phasediagram}~(lower panel).
	
\section{Robustness}
\label{sec:robustness}
In this section we investigate the influence of possible experimental imperfections on values of the retrieved Chern numbers. As an example we choose the Harper-Hofstadter Hamiltonian \eqref{eq:hhHamiltonian}  with the flux $\phi=1/3$ and the finite lattice consisting of $7\times7$ elementary magnetic cells ($7\times21$ lattice sites). All presented quantities are averaged over 90 phase retrieval runs which correspond to different randomly chosen initial states. Percentage of discarded worst (according to error $\varepsilon$, Eq. \eqref{epsilon_error}) retrieval results is either 10\% or 90\%. The error bars are the standard deviations of the averaged values. 

	\begin{figure*}[bt]

\includegraphics[width=\linewidth]{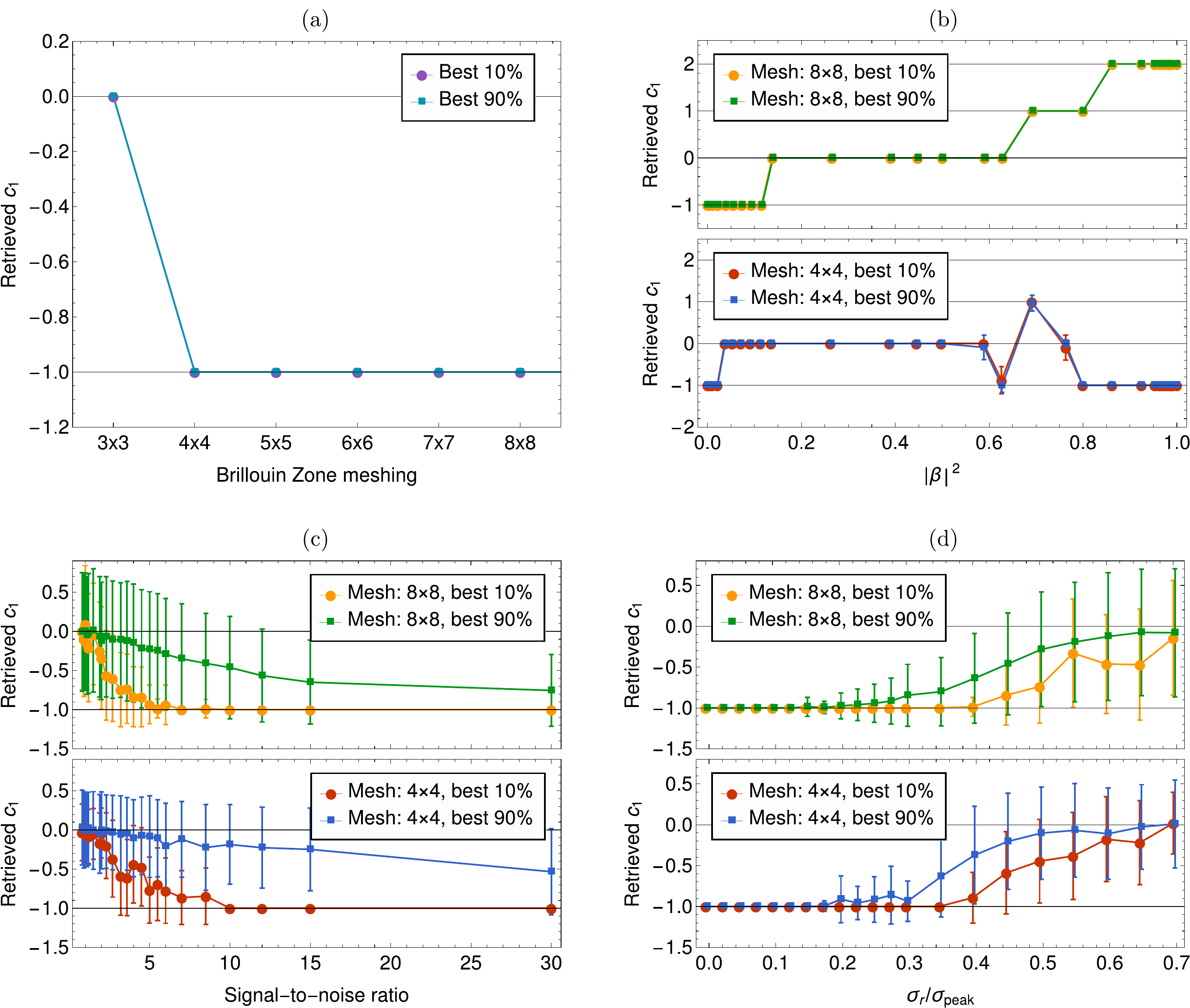}

	\caption{
		Analysis of the influence of experimental imperfections on the retrieved Chern numbers. All presented results are related to the lowest energy band of the Harper-Hofstadter model with the flux $\phi=1/3$. We consider a finite lattice consisting of $7\times7$ elementary magnetic cells  ($7\times21$ lattice sites), cf. Fig.~\ref{fig:hofstadter}. Panel (a): average values of the obtained Chern number $c_1$ of the lowest band as a function of the 
		Brillouin zone meshing. It turns out that is sufficient to  
		discretize the first BZ with a $4\times4$ only in order to obtain the correct value of $c_1$.   Panel~(b): impact of the excitation of the system to the second energy band.  In Harper-Hofstadter model with $q=3$ bands, the lowest energy band corresponds to $c_1=-1$ while the Chern number of the second band is $c_2=2$ (see Fig.~\ref{fig:hofstadter}). For $8\times 8$ meshing, when the occupation of the second band exceeds $|\beta|^2\approx0.12$, the obtained Chern number of the lowest band becomes incorrect, i.e. it switches from $c_1=-1$ to 0. When $|\beta|^2\gtrsim0.86$, the system is actually in the second band and the value of the obtained Chern number equals 2 as expected. For a $4\times4$ mesh, this limit is much smaller: $|\beta|^2\leq0.02$, and the Chern number of the higher band is not correctly reproduced. 
		Panel~(c): dependence of average values of the Chern number on the signal-to-noise ratio. In order to successfully reproduce the Chern number, the signal-to-noise has to be greater than about 5.5. It is evident that discarding more retrieval results reduces the limitation.
		Panel~(d): average values of the Chern number for different resolutions of an experimental imaging system. Finite resolution is simulated by convolution of the atomic density after time-of-flight with the Gaussian function of width $\sigma_r$. Horizontal axis shows $\sigma_r$ in units of the width $\sigma_{\text{peak}}$ of the highest Bragg peak observed in the atomic density after TOF. If $\sigma_r /\sigma_{\text{peak}} \lesssim 0.4$,  the retrieved Chern number is correct.
	}\label{fig:imperfections}
\end{figure*}
	\subsection{Number of points chosen in the first Brillouin Zone}
	\label{sec:npts}
	We have tested how densely one has to probe the first BZ in order to get the proper value of the Chern number $c_1$ corresponding to the lowest energy band in Fig.~\ref{fig:hofstadter}~(left panel). Figure~\ref{fig:imperfections}(a) indicates that it is sufficient to perform the $4\times4$ mesh discretization of the Brillouin zone and the retrieved Chern number is correct. It also demonstrates how powerful the FHS method is. In order to make sure that a Chern number is retrieved correctly, an experiment should be performed again with different discretization  of the BZ.

\subsection{Excitations to the second band}\label{subsec:excitations}
Experimental preparation of an eigenstate from the lowest energy band is usually not perfect and contributions from the higher bands can be expected. In this subsection we analyze contamination of eigenstates of the lowest (first) band $\psi_{\bm k}^{[1]}$ by eigenstates from the second band $\psi_{\bm k}^{[2]}$,
\be
\psi_{\bm k}=\alpha\; \psi_{\bm k}^{[1]} + \beta\; \psi_{\bm k}^{[2]}, \qquad\left|\alpha\right|^2 + \left|\beta\right|^2 = 1,
\ee	
and its influence on the determination of the Chern number $c_1$, which would estimate the worst case scenario for the Landau-Zener transition, see Sec.~\ref{subsec:prep}.

We have applied our method for different populations $|\beta|^2$ of the second band and the results are presented in Fig.~\ref{fig:imperfections}(b).
We conclude that  $\left|\beta\right|^2 \lesssim 0.12$ allows  for the correct retrieval of the Chern number $c_1$ in case of $8 \times 8$ mesh. A similar, symmetric result applies to the Chern number of the second band:  to obtain successfully $c_2$ we require $\left|\beta\right|^2 \gtrsim 0.86$. If one takes only $4 \times 4$ mesh, $\beta$ must satisfy $\left|\beta\right|^2 \lesssim 0.02$ to recover the Chern number of the lowest band. Note that the mesh size in FHS method must be increased  with the absolute value of the Chern number  \cite{FHS:2005}, and therefore in the case of the $4 \times 4$ mesh it is not enough to recover a correct Chern number $c_2=2$. Although we find that the higher  mesh gives a better critical $\left|\beta\right|^2$, at some point the undesired occupancies of other bands will always spoil the results. Therefore, in Sec.~\ref{subsec:prep} we propose a method to minimize the excitations to higher bands.

	\subsection{Background noise}
In the experiment, background noise will affect the atomic density measurements. Let us define the signal strength $A$ as the average value of $\left|\tilde{\psi}_{\bm{k}}(\bm q)\right|^2$ calculated in the first BZ. The signal-to-noise ratio reads $\text{SNR} = A/\sigma_n$, where $\sigma_n$ is the standard deviation of Gaussian white noise whose absolute values are added to each point $\bm q$ of the atomic density image. The results of the retrieved Chern number versus SNR are presented in Fig.~\ref{fig:imperfections}(c). The minimal $\text{SNR}$ that allows for the successful retrieval of the Chern number is about 5.5 for a $8\times8$ mesh after discarding about 90 percent of the worst retrievals. It is also important to note that experimental noise can be reduced either by repeating the experiment and averaging the recorded density profiles over separate realizations, or by applying noise removal algorithms \cite{Vijaykumar2010,Lenzen2013,niu2018}.

	\subsection{Resolution of experimental imaging system}
	In order to check how the results are sensitive to the resolution of the imaging system, we convolve the original atomic density after TOF, $|\psi_{\bm k}(\bm q)|^2$,  with the Gaussian profile of width $\sigma_{r}$. In Fig.~\ref{fig:imperfections}(d) we can see how the average value of the Chern number $c_1$ depends on the ratio $\sigma_{r}/\sigma_{\text{peak}}$, where $\sigma_{\text{peak}}$ is the width of the Gaussian fit to the highest Bragg peak that can be observed in the atomic density, $|\psi_{\bm k}(\bm q)|^2$, after TOF. The minimal resolution that guarantees the correct value of the Chern number is 
	$\sigma_{r}/\sigma_{\text{peak}} \approx 0.4$, which also requires a $8\times8$ mesh after discarding about 90\% of the worst retrieval results.

\section{Summary}
\label{sec:concl}
We have proposed a method for determination of the topological invariants of two-dimensional Chern insulators with the help of ultra-cold bosonic atoms in optical lattice potentials. The method relies on a sequence of experiments where a Bose-Einstein condensate is prepared in different eigenstates of a given energy band. In each experiment, an atomic density after time-of-flight is measured. Because the time-of-flight is actually the Fourier transform of the initial condensate wavefunction of atoms prepared in a finite optical lattice, a phase retrieval algorithm can be applied in order to obtain the phase of the wavefunction. The full knowledge of eigenstates of a given band allows one to calculate the Chern number characterizing the band.

We illustrate the application of the method with two examples: the Harper-Hofstadter model and the Haldane model on a brick wall lattice. It turns out that it is sufficient to retrieve a small number of eigenstates of a given band, i.e. to discretize coarsely the first Brillouin zone, in order to determine the Chern number. An experimental sequence that allows one to avoid population of neighboring bands, during the preparation of the system in a topological phase, is presented. We also analyze robustness of the method and its resistance to experimental imperfections.

\section*{ACKNOWLEDGMENTS}

Support of Talenty and Grand Scholarship Programs (T.S.), the National Science Centre, Poland via Projects No.~2016/21/B/ST2/01095 (T.S.), No.~2016/21/B/ST2/01086 (A.K.) and QuantERA programme No.~2017/25/Z/ST2/03027 (K.S.) is acknowledged.
\appendix

\section{Fukui-Hatsugai-Suzuki Method}
\label{app:FHS}

 Assume the  2D system on a square lattice that is invariant under discrete space translations $x\rightarrow x+q_x$ and $y\rightarrow y+q_y$, where $q_x$, $q_y$ are integer  multiples of the lattice constant $a=1$. Hence, the system can be described completely by a $q_xq_y \times q_xq_y$ Hamiltonian matrix  $\mathcal{H}\left(\bm{k}\right)$  in a reduced Brillouin zone $\bm k \in \left(-\pi/q_x, \pi/q_x\right] \times \left(-\pi/q_x,\pi/q_x \right]$.  
 Assume that, for each $\bm k$, the Hamiltonian $\mathcal{H}\left(\bm{k}\right)$ has non-degenerate eigenvalues. Then, the solutions of the Schr\"{o}dinger equation

\begin{equation}
		\mathcal{H}\left(\bm{k}\right) \bm{u}^{[n]}\left(\bm{k}\right) = E^{[n]}  \left(\bm{k}\right)\bm{u}^{[n]}  \left(\bm{k}\right),
\end{equation}
 describe separate energy bands labeled by $n=1,\ldots,q_xq_y$.  Let us take a set of discrete points $\bm{k}_l$ $\left(l=1, \dots, N_x N_y\right)$ in the first BZ
\begin{eqnarray*}
\bm{k}_l &=& \left(k_{l_1}, k_{l_2} \right), 
\end{eqnarray*}
with
\begin{eqnarray*}
k_{l_\mu} &=& \frac{2 \pi l_\mu}{q_{\mu} N_\mu}, \qquad l_\mu = 0, 1, \dots N_\mu-1,
\end{eqnarray*}
where $\mu = x,y$.  We will call $\hat{\mu}$ the vector of the length $\delta k_\mu = 2\pi /\left(q_\mu N_\mu\right)$ in the direction $\mu$. The $U(1)$ linking variables of the $n$-th band are defined as
\begin{equation}
U^{[n]}_\mu\left(\bm{k}_l\right) := \bm{u}^{[n] \dagger}\left(\bm{k}_l\right)\bm{u}^{[n]}\left(\bm{k}_l + \hat{\mu}\right)/\mathcal{N}^{[n]}_\mu\left(\bm{k}_l\right),
\end{equation}
with $\mathcal{N}^{[n]}_\mu\left(\bm{k}_l\right) = \left| \bm{u}^{[n]\dagger}\left(\bm{k}_l\right)\bm{u}^{[n]}\left(\bm{k}_l + \hat{\mu}\right)\right|.$

The field strength $\tilde{F}^{[n]}_{xy}\left(\bm{k}_l\right)$ takes a manifestly gauge-invariant form
\begin{align}
&\tilde{F}^{[n]}_{xy}\left(\bm{k}_l\right) := \ln\frac{U^{[n]}_x\left(\bm{k}_l\right) U^{[n]}_y\left(\bm{k}_l + \hat{x}\right)}{  U^{[n]}_x\left(\bm{k}_l + \hat{y}\right) U^{[n]}_y\left(\bm{k}_l\right)}, \\\nonumber &-\pi < \frac{1}{i}\tilde{F}^{[n]}_{xy}\left(\bm{k}_l\right) \leq \pi.
\end{align}

Finally, the Chern number reads
\begin{equation}
c_n = \frac{1}{2\pi i}\sum_{l} \tilde{F}^{[n]}_{xy}\left(\bm{k}_l\right).
\end{equation}
Even for coarsely discretized BZ's this algorithm gives accurate values of the Chern numbers (see Sec. \ref{sec:npts} or ref. \cite{FHS:2005}).

\begin{figure*}[tbh]
	
	\includegraphics[width=.85\textwidth]{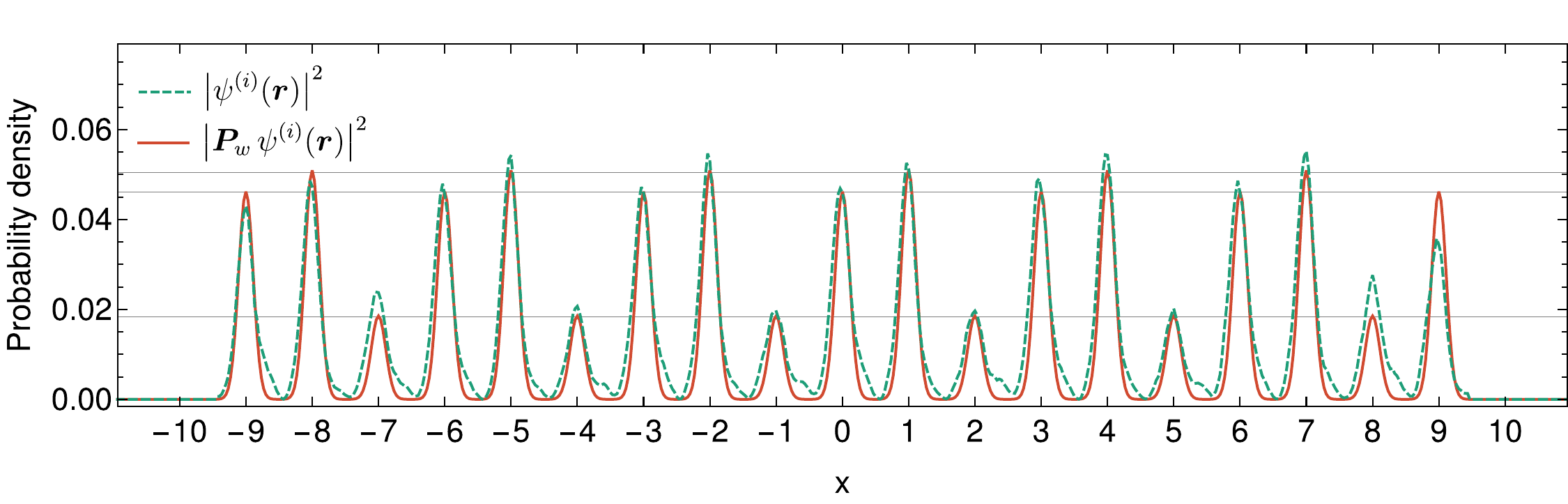}
	\caption{Illustration of the result of the projection $\bm{P}_w = \bm{P}_{w_2} \circ \bm{P}_{w_1}$ that is used in the phase retrieval algorithm. The Harper-Hofstadter model with the flux $\phi=1/3$ is considered. The plot shows a cut of the probability density along the $x$ direction before (dashed line) and after (solid line) the projection. The projection reestablishes the translational symmetry of the system.
		 Horizontal lines help to see that due to the projection, the probability densities in all sublattice sites become equal.}
	\label{fig:Pwannier}
\end{figure*}

\section{Phase retrieval algorithm and its optimization}
\label{app:PR}
Phase retrieval algorithms iteratively seek for a solution $\psi\left(\bm{r}\right)$ in the object space, provided the modulus of its Fourier transform $\left|\tilde{\psi}_{\bm{k}}\left(\bm{k}\right)\right|$ and support S (area where $\psi\left(\bm{r}\right) \neq 0$)  are known. The simplest version of the algorithm, called \emph{error reduction} (ER), is described in Sec. \ref{subsec:tof}. Fienup proves \cite{Fienup:1978} that at each iteration, the retrieval error, defined as 
\begin{equation}
	\varepsilon = \int \mathrm{d}^2 \bm{q}~\left( \left|\tilde{\psi}^{(i)}\left(\bm{q}\right)\right| - \left|\tilde{\psi}_{\bm{k}}\left(\bm{q}\right)\right| \right)^2,
	\label{errorapp}
\end{equation}
decreases. Stagnation of this algorithm in local minima is, however, likely to occur \cite{Fienup1982,Marchesini2007}. Several approaches have been proposed to solve this problem \cite{Marchesini2007}. One example is the \emph{hybrid input-output} (HIO) algorithm based on nonlinear feedback control theory
\cite{Fienup1982}. It is very similar to the ER algorithm, the only change is the step \hyperref[ER:iv]{(iv)} described in Sec.~\ref{subsec:tof}. The part of $\psi^{(i)}\left(\bm{r}\right)$ that lies outside the support is not set to zero but instead to $\left(1 - \eta \bm{P}_m\right) \psi^{(i)}\left(\bm{r}\right)$, where the operator \mbox{$\bm{P}_m$} (described in steps \hyperref[ER:i]{(i)-(ii)} in  Sec.~\ref{subsec:tof}) is the projection on the set of functions with the modulus $\left|\tilde{\psi}_{\bm{k}}\left(\bm{k}\right)\right|$ and $0<\eta<1$ is the feedback parameter, usually set to $0.7-0.9$. In most cases, a combination of the HIO and ER methods, e.g. 20 iterations of HIO and 1 iteration of ER algorithms, repeated in cycle, gives the best results. Since the HIO method does not guarantee the decrease of the error $\varepsilon$, the last few (30-50) iterations, should consist of the pure ER algorithm.
\paragraph*{Support}
If we want to recover $N$ \textit{complex} numbers $\psi_{\bm{k}}\left(\bm{r}\right)$ within support, we need at least $2N$ \textit{real} numbers $\left|\tilde{\psi}_{\bm{k}}\left(\bm{q}\right)\right|$. This gives a constraint on the area of the support which must not be less than $50\%$ the area of the whole table of $\psi_{\bm{k}}\left(\bm{r}\right)$. In our case, the support occupies only $22.5\%$ of the whole table which increases the rate of convergence.
If the support is symmetric with respect to rotation by 180 degrees around some point $\bm{r}_0$ in space (e.g., the support is a rectangle or a circle), the fact that $\psi_{\bm{k}}\left(\bm{r}-\bm{r}_0\right)$ and $\psi^*_{\bm{k}}\left(-\left(\bm{r}-\bm{r}_0\right)\right)$ have the same modulus of the Fourier transform causes an ambiguity. The algorithm will converge to any of the two solutions with equal probability and in some cases it will stagnate at their superposition \cite{Fienup:86}. The only other nonuniqness can appear if and only if $\psi_{\bm k}(\bm r)$ can be written as a convolution of two non-central symmetric functions \cite{Barakat1984}. Therefore, in our simulations we choose a trapezoidal support with the ratio 4/5 of its bases which corresponds to a hard-wall box potential of this shape. Fluctuations of the size of an atomic cloud in a trap  result in  changes of the width of the Bragg peaks in the momentum distribution. The latter are not dangerous in the determination of Chern numbers. We also stress that if the size of the cloud is fluctuating one must set a support that is slightly larger than the average size of the cloud. This way one does not unintentionally "cut" the solution in real space.

\paragraph*{Optimization}
\begin{figure}[htpb]
	
	\includegraphics[width=\columnwidth]{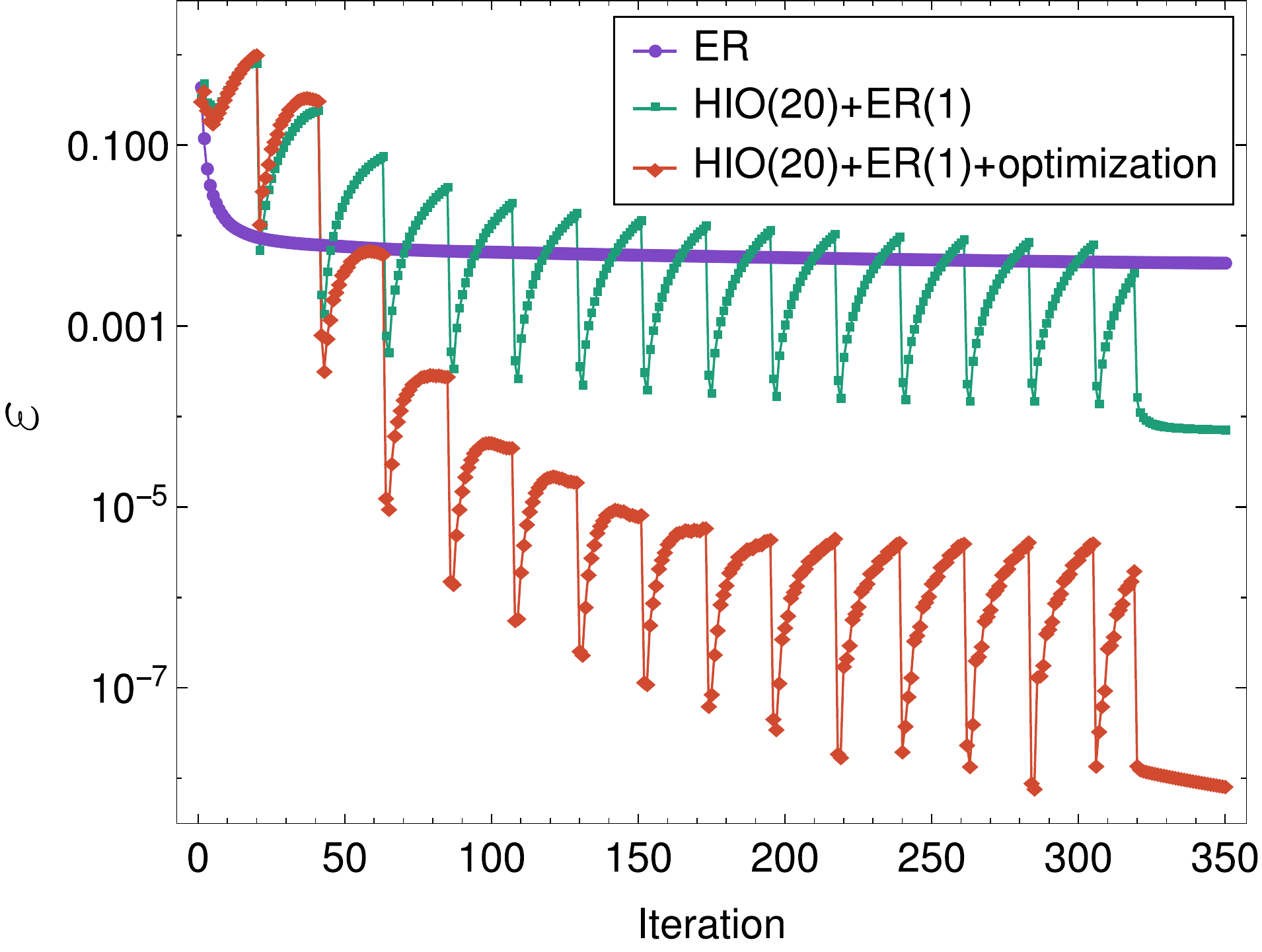}
	\caption{Comparison of different versions of the phase retrieval algorithm. The combination of the HIO and ER methods leads to a smaller error $\varepsilon$, see Eq.~(\ref{errorapp}), than the ER method alone but to a much larger error than in the case when the projection $\bm{P}_w = \bm{P}_{w_2} \circ \bm{P}_{w_1}$ is applied every third iteration, see the discussion in the text.
		{\it Fringes} correspond to one cycle of the HIO(20)+ER(1)=21 iterations. The last 30 iterations correspond to the pure ER method which allows one to reduce the error at the end of the retrieval process. }
	\label{fig:prmod}
\end{figure}
If additional information about $\psi\left(\bm{r}\right)$ is known, it can be used to speed up the algorithm convergence. For example if the geometry of an optical lattice and the number of lattice sites can be estimated in the experiment, we know all information about an eigenstate of the system presented in Eq.~\eqref{eq:wf} except the factors $e^{i \bm{k} \cdot \bm{r}_{\ell \alpha}} u_{\alpha}$. We use this information as follows. 

Define the projection $\bm{P}_{w_1}\psi^{(i)}$ of a current estimate of the desired solution on the Wannier state basis, 
\begin{equation}
\bm{P}_{w_1} \psi^{(i)}\left(\bm{r}\right) := \frac{1}{\mathcal{N}}\sum_{\ell,\alpha} v_{\ell \alpha}^{(i)} w\left(\bm{r}-\bm{r}_{\ell \alpha}\right),
\end{equation}
where $\ell=1, \ldots, n_{\text{cells}}$ is the index of an elementary cell, $\alpha~=~1,\ldots,q$ is the index of a lattice site within an elementary cell, $\mathcal{N}$ is the normalization factor and 
\begin{equation}
v_{\ell \alpha}^{(i)} = \int \mathrm{d}^2 \bm{r}~w^*\left(\bm{r}-\bm{r}_{\ell s}\right) \psi^{(i)}\left(\bm{r}\right).
\end{equation}
If $\psi^{(i)}\left(\bm{r}\right)$ is identical with the desired solution, then $v_{\ell \alpha}^{(i)} \equiv e^{i \bm{k} \cdot \bm{r}_{\ell \alpha}} u_{\alpha}$, hence $|v_{\ell \alpha}^{(i)}|$ should not depend on $\ell$ and we impose this condition in the iterative process. We define the next projection $\bm{P}_{w_2}$,
\begin{equation}
\bm{P}_{w_2}\left(\bm{P}_{w_1}  \psi^{(i)}\left(\bm{r}\right)\right) := \sum_{\ell,\alpha}  \abs{v_{\alpha}^{(i)}}_{\text{rms}} e^{i \Arg v_{\ell \alpha}^{(i)}}w\left(\bm{r}-\bm{r}_{\ell \alpha}\right),
\end{equation}
where
\begin{equation*}
\abs{v_{\alpha}^{(i)}}_{\text{rms}}^2 = \frac{1}{n_{\text{cells}}} \sum_{\ell }\abs{v_{\ell \alpha}^{(i)}}^2
\end{equation*}
is the mean occupation of the sublattice site $\alpha$. This operation ensures that occupations of the same sublattice sites in all elementary cells are the same (see Fig.~\ref{fig:Pwannier} for clarification). The complete projection  
\begin{equation}
\bm{P}_w = \bm{P}_{w_2} \circ \bm{P}_{w_1},
\end{equation}
is performed every 3 iterations of the phase retrieval algorithm.  The effect of our optimization is clearly visible in Fig.~\ref{fig:prmod} --- the final error (\ref{errorapp}) is about 4 orders of magnitude smaller than without the optimization (see also comprehensive phase retrieval software libraries \cite{chandra2017phasepack}).

%\bibliography{ref_10_2018.bib} 
\input{article_v15.bbl}

\end{document}

%% file: article_v15.bbl

%% file: article_v15.bbl
\begin{thebibliography}{78}
\bibitem [{cSe()}]{cSecond}%
  \BibitemOpen
  \href@noop {} {}\bibinfo {note} {Note that we consider topological properties
  of non-interacting systems only. If there are non-negligible interactions
  between atoms they can be turned off during the time-of-flight expansion by
  using Feshbach resonances}\BibitemShut {NoStop}%
\end{thebibliography}
